*Direct measurement of non-equilibrium phonon occupations in femtosecond laser heated Au films*

T. Chase,[1,2] M. Trigo,[1] A. H. Reid[1], R. Li,[1] T. Vecchione[1], X. Shen[1], S. Weathersby[1], R. Coffee[1], N. Hartmann[1], D.A. Reis[1], X.J. Wang[1] and H. A. Dürr[1]

[1]SLAC National Accelerator Laboratory, 2575 Sand Hill Road, Menlo Park, California 94025, USA

[2]Department of Applied Physics, Stanford University, Stanford, California 94305, USA

We use ultrafast electron diffraction to detect the temporal evolution of phonon populations in femtosecond laser-excited ultrathin single-crystalline gold films. From the time-dependence of the Debye-Waller factor we extract a 4.7 ps time-constant for the increase in mean-square atomic displacements. We show from the increase of the diffuse scattering intensity that the population of phonon modes near the X and K points in the Au fcc Brillouin zone grows with timescales of 2.3 and 2.9 ps, respectively, faster than the Debye-Waller average. We find that thermalization continues within the initially non-equilibrium phonon distribution after 10 ps. The observed momentum dependent timescale of phonon populations is in contrast to what is usually predicted in a two-temperature model.

Understanding of the scattering processes due to the interactions between electrons and lattice vibrations is of importance in a wide range of condensed matter phenomena such as superconductivity [1], electronic and spin transport [2, 3], ultrafast demagnetization [4] and laser induced phase transitions [5, 6]. The availability of fs lasers allowed the detailed study of electronic scattering, thermalization and electron-phonon energy transfer in real time [7-11]. Following the rapid thermalization of the electronic system within several 100 fs of laser excitation, the coupled electron-phonon systems are usually treated by the so-called two-temperature model that assigns different temperatures to electrons and lattice [10, 11]. Time-resolved X-ray scattering experiments on single-crystal semiconductors found evidence that the lattice remains out of equilibrium for hundreds of picoseconds, and thus a lattice temperature is not well-defined [12, 13]. More recent experiments at free electron lasers [14] show that the sub-picoseconds dynamics is highly non-thermodynamic.

Here we use ultrafast electron diffraction to probe the temporal evolution of both Bragg peaks and the diffuse scattering after fs laser excitation of a 20 nm Au film. This allows us to determine the temporal evolution of phonon populations at wavevectors ranging from the Au Brillouin zone center to the zone boundary. We observe a wavevector-dependent increase of the diffuse scattering intensity. Near the X point of the Au Brillouin zone, the time constant of 2.3 ps is faster than the 4.7 ps rise time of the average mean-square displacements obtained from the Debye-Waller attenuation of Bragg peaks. Our results indicate that longitudinal acoustic (LA) polarizations. phonon modes populated faster than transverse acoustic (TA) modes in agreement with predicted electron-phonon coupling strengths [15]. The non-equilibrium behavior of the phonon populations for 10 ps and above demonstrates that the energy transfer from the laser-heated electrons does not instantaneously result in an equilibrated phonon system that can be described by a single lattice temperature as assumed in the two-temperature model.

Fig. 1 shows the schematic layout of the experiments. 3.3 MeV electron pulses of 200 fs (FWHM) duration from the SLAC ultrafast electron diffraction (UED) facility [16] were transmitted through Au films at normal incidence and the scattering pattern was detected with a phosphor screen based EMCCD detector. Contrary to previous UED studies that employed poly-crystalline samples [17], we used single-crystalline gold films of 20 nm thickness (SPI Inc.) supported by a $Si_3N_4$ membrane of similar thickness floated onto standard 100 μm wide TEM grids. We selected Au samples with the highest degree of crystallinity resulting in the sharpest Bragg peaks with the surface normal along the [001] lattice direction. Laser pump pulses of 800 nm central wavelength and 40 fs (FWHM) duration were coupled nearly collinearly with the electron beam by an optical mirror that enabled transmission of the electron beam though a hole in its middle. This resulted in laser pump and electron probe spot sizes at the sample of 0.4 mm and 1.6 mm diameter, respectively [16]. The laser pump fluence used for the experiments in this paper was 3.2 mJ/cm$^2$, low-enough to prevent any laser-induced permanent changes to the Au sample. We note that this fluence was so far below the melting threshold in ref. [17] that we also did not have to take any changes in the phonon frequencies into account. The pump-probe time delay was scanned over 50 ps after time zero. The pumped and un-pumped diffraction patterns were taken in an alternating sequence. Each pattern was accumulation over 120 electron pulses on the detector. At each pump - probe time delay a total of 800 pumped and 800 un-pumped diffraction patterns were taken. We used the total integrated intensity of the un-pumped diffraction patterns to monitor and correct for long-term electron beam intensity drift. The scattering patterns were corrected for background due to the readout noise of the CCD camera as determined via measurements without any beam. We extracted the Debye-Waller attenuation of the (400) and (220)-type Bragg peaks following ref. [18] by dividing through the (200) Bragg peak intensity that showed a much smaller intensity variation owing to the lower-order reciprocal lattice vector for this reflection. The time-resolved diffuse scattering signal was obtained by subtracting pumped scattering patterns after time zero and a reference pattern taken at -2 ps time delay. Because of the low repetition rate (120Hz) there were no pump accumulative effects at -2 ps.

Time-resolved scattering patterns were measured for pump - probe time delays up to 50 ps. Figs 2 (a-c) shows representative scattering patterns obtained for selected time delays. Initially, 2 ps after laser excitation the diffuse intensity is low but evenly distributed over most of the reciprocal space (a). The diffuse intensity becomes stronger and more anisotropic as the time delay increases, (b) – (c). There is strong diffuse scattering close to the Bragg peaks (dark blue disks in Fig. 2), which continues to grow up to 50 ps. In addition to the intensity buildup near the Bragg peaks, faint streaks of intensity start to appear in-between them. Interestingly these streaks saturate at different times depending on position between the (220) and (400)-type Bragg peaks at which the measurements are taken. We will analyze this behavior further in Fig. 3.

To understand the observed anisotropic distribution of the diffuse scattering intensity we performed simulation of the diffuse scattering pattern of Au for thermalized phonons (see Fig. 2d). The phonon frequencies and eigenvectors were computed using density functional perturbation theory as implemented in the Quantum ESPRESSO package [19]. We used the simulation parameters for gold reported in ref. [20]. The intensity of scattered electrons, $I$, was calculated for a thermalized phonon population at 300 K. In a single-scattering process, $I$, is linked to the phonon occupation, $n$, as [13, 21]

$$I(\mathbf{Q}) \propto \sum_j \frac{1}{\omega_j(\mathbf{q})}\left[n_j(\mathbf{q}) + \frac{1}{2}\right]|F_j(\mathbf{Q})|^2 \qquad (1)$$

where $\mathbf{Q}$ is the wavevector exchanged in the scattering process, $\omega_j(\mathbf{q})$ is the frequency of the phonon mode in branch $j$ with reduced wave vector $\mathbf{q}$, $n_j(\mathbf{q})$ is its population and $F_j(\mathbf{Q}) \propto \sum_s \frac{f_s}{\sqrt{m}} e^{-M_S}[\mathbf{Q} \cdot \boldsymbol{\varepsilon}_j] e^{-i\mathbf{K}_Q \cdot \mathbf{r}_s}$ [21]. In $F_j(\mathbf{Q})$, $f_s$, $m_S$, and $M_S$ are the atomic scattering factor, the mass, and the Debye-Waller factor, respectively, of atom $s$ at position $\mathbf{r}_s$, $\boldsymbol{\varepsilon}_j$ is the phonon polarization vector, and $\mathbf{K}_Q$ is the closest reciprocal-lattice vector to $\mathbf{Q}$, i.e., $\mathbf{Q} = \mathbf{q} + \mathbf{K}_Q$. It is important to point out the scattering wavevector, *Q*, is measured with respect to the center of the patterns in Fig. 2 while the phonon wavevector, *q*, is measured with respect to the respective Bragg peak positions corresponding to the Brillouin zone centers, Γ (see Fig. 2d).

In the following we will analyze our diffuse scattering data for phonon wavevectors between the (220) and (400)-type Bragg peaks corresponding to the Γ-K-X-K-Γ high-symmetry line (see Fig. 2a) of the Au fcc Brillouin zone using eq. (1). Fig. 3a shows the calculated phonon dispersion, $\omega_j(\mathbf{q})$, along this line. The modes are separated into branches with TA and LA polarizations. We note that the factor $\mathbf{Q} \cdot \boldsymbol{\varepsilon}_j$ in $F_j(\mathbf{Q})$ of eq. (1) means that the intensity at a particular *Q* is dominated by phonons polarized along *Q*. Our experimental geometry is not sensitive to TA modes with polarization along the [001] surface normal (dashed line in Fig. 3a). Instead we are sensitive to TA (solid blue line in Fig. 3a) modes along [110] at Γ(220) (see Fig. 2a) and also increasingly sensitive to LA modes (solid red line in Fig. 3a) when moving along the Γ-K-X-K-Γ line from (220) towards (400) Bragg peaks. At Γ(400) LA and TA modes are probed with equal weight.

Fig. 3b shows the temporal evolution of the change in scattering intensity along the Γ-K-X-K-Γ line between (220) and (400) Bragg peaks. In order to improve statistics we averaged the intensities inside a band of wavectors with a width of 0.2 reciprocal lattice units (rlu) perpendicular to this high-symmetry direction (see Fig 2a) as well as over all four equivalent vertical high-symmetry lines between (220) and (400) Bragg peaks. The temporal evolution of the diffuse scattering intensity integrated for the shaded regions in Fig. 3b is plotted in Fig. 3d (symbols). The widths of the individual regions 1-6 in Fig. 3b was adjusted to ensure roughly similar statistics for each region after averaging. The symbols in Fig. 3c show the Debye-Waller attenuation of the (220) and (400)-type Bragg peaks for comparison. The temporal behavior of the Debye-Waller factor and the diffuse scattering are well described by a single exponential (solid-lines in Figs 3c and 3d.). The obtained time constants and the corresponding $\chi^2$ errors are summarized in table 1.

The Bragg peak intensity decreases due to Debye-Waller effects (dips in all curves of Fig. 3b near Γ) with a typical time constant of 4.7 ± 0.3 ps (Fig. 3c and Tab. 1). In the following we show that the diffuse scattering results in Fig. 3b allow us a far more detailed insight into the energy transfer form the laser-heated electronic system to the lattice. Figs 3b and 3d show a clear allover increase of the diffuse scattering intensity along Γ-K-X-K-Γ with increasing time delay for the first 8 ps. At longer time delays the intensity near X (lowest intensity curve in Fig. 3d) is saturating near 8 ps and even slightly decreases when approaching 50 ps. Approaching the (220) and (400) Bragg peaks the diffuse intensity is seen to saturate at much later times. This is most pronounced in regions 1, 2 and 6 (the three highest intensity curves in Fig. 3d) where saturation is only reached near 50 ps. Fig. 3b (colored bars) and Tab. 1 display the systematic change of the time constants along Γ-K-X-K-Γ. The increase of the phonon population is fastest near the X-point of the Brillouin zone and continuously changes over K to Γ towards longer timescales. Interestingly the increase in timescale is smaller towards Γ(220) compared to approaching the Γ(400) point. Near the two nominally equivalent Brillouin zone centers in regions 1 and 6 we obtain time scales of 4.3 ± 0.2 ps and 3.3 ± 0.2 ps, respectively. We assign this difference to the different weight with which LA and TA modes are probed. Near Γ(220) our experiment is mainly sensitive to the low-frequency TA mode shown in Fig. 3a. The observed timescale of 4.3 ps can therefore be assigned largely to the increase to the low-frequency TA mode population. The shorter timescale observed near Γ(400) can then be assigned to the additional sensitivity to the LA phonon mode at this phonon wavevector which becomes populated faster. Near the X-point (region 3) we cannot separate LA and TA mode contributions. However, our data are consistent with a faster phonon population increase due to stronger electron-phonon coupling near K-X-K than for the rest of the probed reciprocal space volume in agreement with phonon lifetime calculation performed for thermal equilibrium [Tang 2011].

While the mean square displacement increase in Fig. 3c saturates after 10 ps there is still significant change in the phonon population especially close to the Brillouin zone center (regions 1 and 6 in Fig. 3b). On such timescales we can neglect further energy transfer from the laser-heated electrons [22] but we remain sensitive to the thermalization of the phonon system due to anharmonic phonon-phonon scattering. This leads to the intensity distribution at 50 ps which is nearly identical to that expected in thermal equilibrium (Fig. 2d). We point out that a detailed study of the thermalizing phonon system would require probing all phonon modes throughout the whole Au Brillouin zone which is beyond the scope of the present manuscript.

In conclusion, we probed the non-equilibrium phonon dynamics in 20 nm thin Au films following fs optical laser excitation. Phonon wavevectors in multiple Brillouin zones can be probed simultaneously using ultrafast electron diffraction from single-crystalline samples. We showed that the phonon population dynamics along the Γ-K-X high-symmetry direction is faster, ranging between 2.3 to 4.3 ps, than the average lattice disordering rate obtained from Debye-Waller measurement of increasing average mean-square displacements (4.7 ps). In addition, LA modes are more quickly populated than lower-energy TA modes. We also observe that it takes ~ up to 50 ps for the lattice to show a thermal phonon distribution. This establishes the usefulness of ultrafast electron diffraction for phonon studies and paves the way for future experiments accessing phonon dynamics throughout the Brillouin zones of more complex materials.

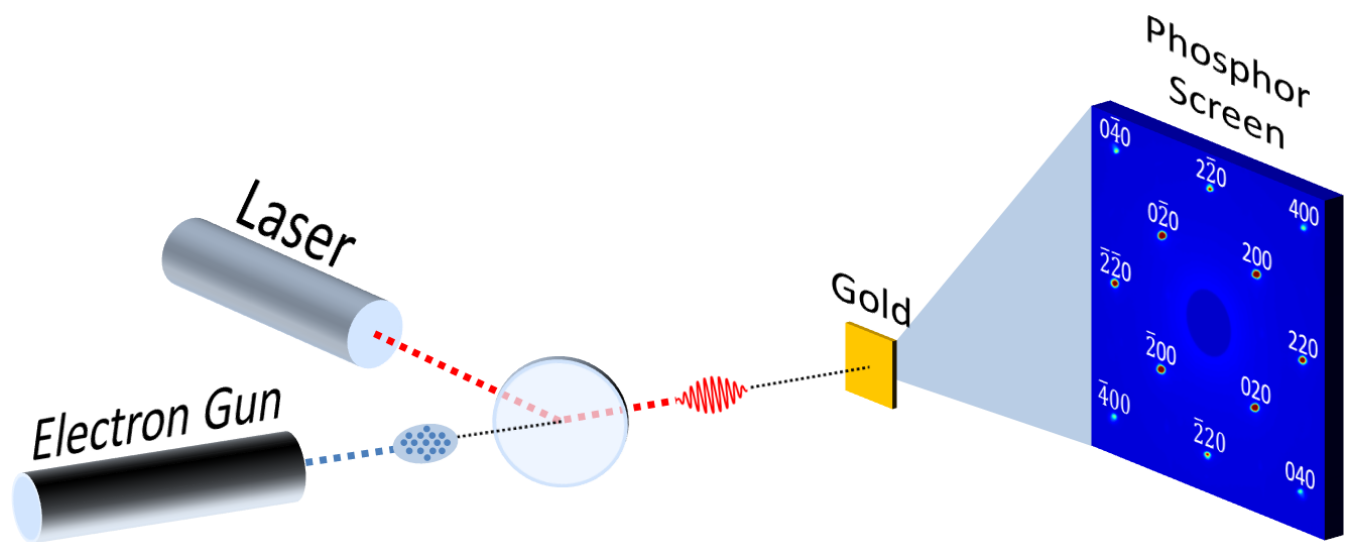

FIG. 1. Schematics of the optical pump - ultrafast electron diffraction probe setup. The shown diffraction pattern shows Bragg peaks of various order as labeled in the figure.

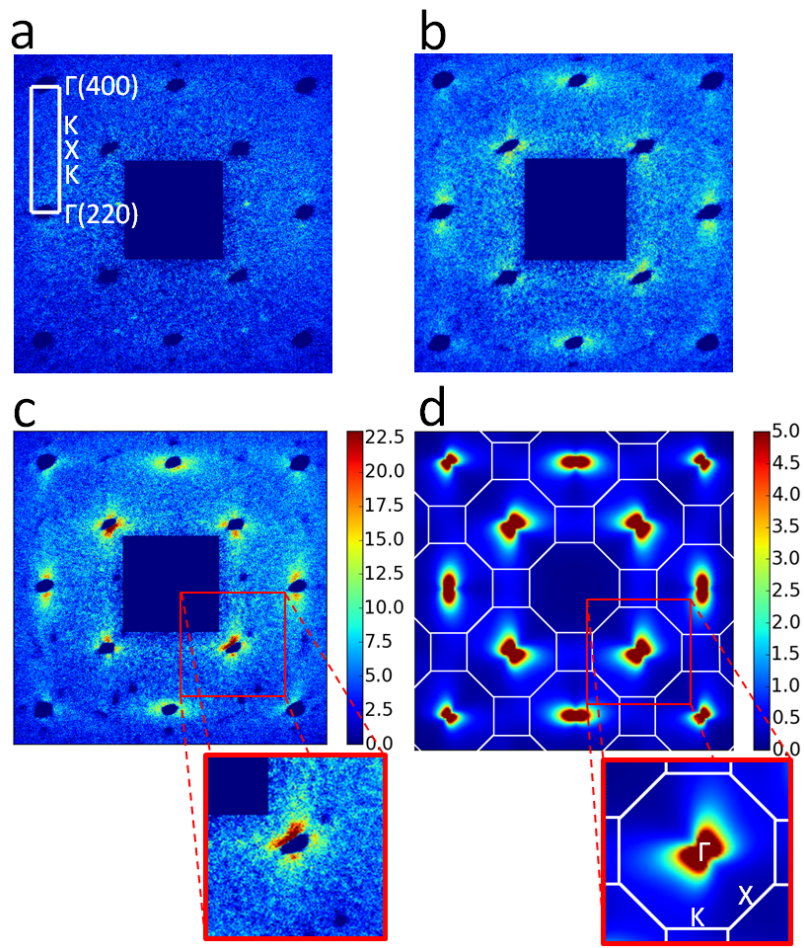

FIG. 2. Differences of diffuse scattering patterns obtained at (a) 2 ps (b) 8 ps and (c) 50 ps pump-probe time delays and their respective static patterns. (d) scattering pattern calculated for a lattice temperature of 300 K with the Brillouin zone boundaries indicated as white lines. The diffuse intensity for wavevectors along Γ(220)-K-X-K-Γ(400) shown in Fig. 3 was evaluated between (400) and (220)-type Bragg peaks, respectively, as indicated in (a).

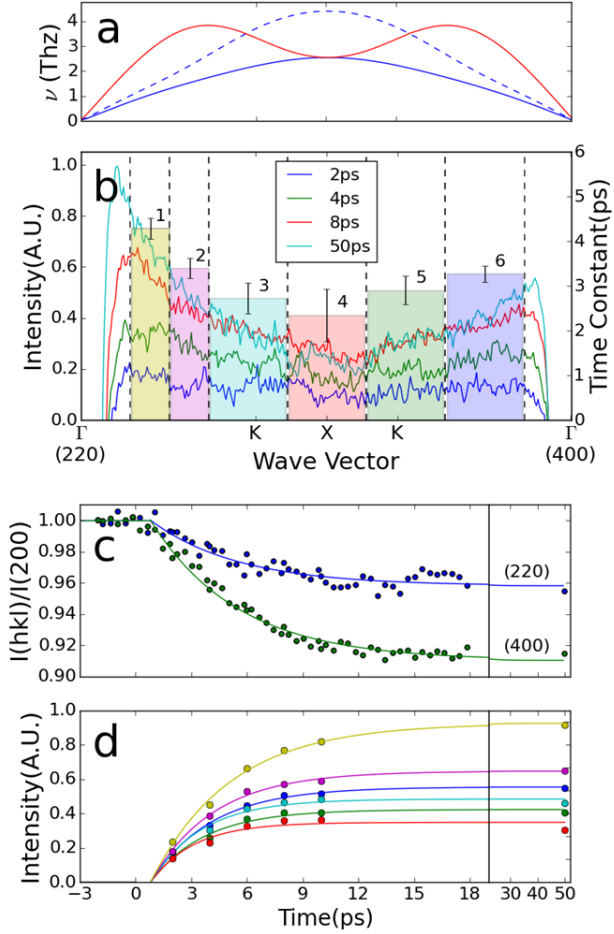

FIG. 3. (a) LA (red) and TA (blue) phonon dispersions along the Γ-K-X-K-Γ line between (400) and (220)-type Bragg peaks shown in Fig. 2a. Solid lines denote phonons with polarization vectors in the plane spanned by the wave vector exchanged in the electron diffraction experiment. (b) time evolution of the scattering intensity along Γ-K-X-K-Γ (lines). The height of the color bars denote the time constants (see (d) and Tab. 1) obtained for the respective wavevector regions. (c) Debye-Waller attenuation of (400) (green lines and symbols) and (220)-type (blue lines and symbols) Bragg peaks vs. pump -probe time delay. A 30 K temperature increase can be deduced from both reflections. (d) scattering intensity in the wavevector regions defined in (b) vs. pump - probe the delay. Symbols in (c) and (d) denote experimental data while lines are exponential fits with the obtained time constants summarized in Tab. 1.

| wave vector region | 1 | 2 | 3 | 4 | 5 | 6 | DW |
|---|---|---|---|---|---|---|---|
| time constant | 4.3 ± 0.2 | 3.4 ± 0.2 | 2.7 ± 0.4 | 2.3 ± 0.6 | 2.9 ± 0.3 | 3.3 ± 0.2 | 4.7 ± 0.3 |

TABLE I. Summary of the time constants extracted from exponential fit to the experimental data shown in Figs. 3c, d. The regions 1-6 are defined by the color bars in Fig. 3b while DW denotes the Debye-Waller data in Fig. 3c.

## ACKNOWLEDGMENTS

The authors would like to thank SLAC management for their continued support. The technical support by SLAC Accelerator Directorate, Technology Innovation Directorate, LCLS Laser Science & Technology division and Test Facilities Department is gratefully acknowledged. This work was supported in part by the U.S. Department of Energy (DOE) Contract No. DE-AC02- 76SF00515, DOE Office of Basic Energy Sciences Scientific User Facilities Division Accelerator and Detector R&D program, the SLAC UED/UEM Initiative Program Development Fund and Laboratory Directed Research and Development funding.